\documentclass[review]{elsarticle}

\usepackage{lineno,hyperref}
\modulolinenumbers[5]

\journal{}

\usepackage{epstopdf}
\usepackage{color}
\usepackage{mathtools}
\usepackage{amsmath}
\usepackage{amsfonts}
\usepackage{amssymb}
\usepackage{enumerate}
\usepackage{subcaption}

\usepackage{scrextend}
\usepackage{amssymb,amsmath,amsthm,enumitem}

\usepackage{listings,inconsolata}

\usepackage{comment}
\usepackage{verbatim}

\usepackage{graphicx}
\graphicspath{ {./images/} }

\usepackage{amsmath}
\usepackage{algorithm}
\usepackage[noend]{algpseudocode}

\usepackage{xcolor}

\usepackage[normalem]{ulem}

\usepackage{tikz}
\usepackage{pifont}

\usepackage{amsfonts}
\usepackage{latexsym}
\usepackage{hyperref}
\usepackage{enumerate}
\usepackage{wasysym}
\usepackage{lipsum}
\usepackage{amssymb}
\usepackage{amsthm}

\usepackage{tikz}
\usepackage{pifont}

\usepackage{color}

\begin{document}

\begin{frontmatter}
\title{Design and Implementation of EEG-Mechatronic System Interface for Computational Intelligence
}




\author[MQU]{Cameron Aume}
\author[QUT]{Shantanu Pal\corref{mycorrespondingauthor}}\cortext[mycorrespondingauthor] {Corresponding author}
\ead{shantanu.pal@qut.edu.au} 
\author[MQU]{Subhas Mukhopadhyay}

\address[MQU]{School of Engineering, Macquarie University, Sydney, NSW 2109, Australia}
\address[QUT]{School of Computer Science, Queensland University of Technology, Brisbane, QLD 4000, Australia}

\begin{abstract}
The devices that can read Electroencephalography (EEG) signals have been widely used for Brain-Computer Interfaces (BCIs). Popularity in the field of BCIs has increased in recent years with the development of several consumer-grade EEG devices that can detect human cognitive states in real-time and deliver feedback to enhance human performance. Several studies are conducted to understand the fundamentals and essential aspects of EEG in BCIs. However, the significant issue of how can consumer-grade EEG devices be used to control mechatronic systems effectively has been given less attention. In this paper, we have designed and implemented an EEG BCI system using the OpenBCI Cyton headset and a user interface running a game. We employ real-world participants to play a game to gather training data that was later put into multiple machine learning models, including a linear discriminant analysis (LDA), k-nearest neighbours (KNN), and a convolutional neural network (CNN). After training the machine learning models, a validation phase of the experiment took place where participants tried to play the same game but without direct control, utilising the outputs of the machine learning models to determine how the game moved. We find that a CNN trained to the specific user playing the game performed with the highest activation accuracy from the machine learning models tested, allowing for future implementation with a mechatronic system.
\end{abstract}


\end{frontmatter}


\section{Introduction}
\label{chap:Introduction}
The interface between humans and computers has long been a challenging problem to solve~\cite{gu2021eeg}. With the development of Electroencephalographic (EEG) headsets, there has been a particular interest in reading brainwaves to develop a Brain-Computer Interface (BCI)~\cite{pal2022brain}. 
~In 1890, Adolf Beck discovered that changing voltage potentials could be measured on the heads of some animals in response to sensory stimulation \cite{beck_1890}. This was one of the first experiments which displayed changing electrical voltages on the head. Over the years, there have been many developments to find ways to detect the electrical activity of the human brain. Towards this, EEG is a safe, painless, and efficient method for humans. EEG reads voltage potentials on the human's scalp. The first EEG experiment was conducted in 1924 when electrical activity of the human brain was measured using non-invasive techniques~\cite{hans_berger}. Since the first developments of EEG, it has been found useful in many diverse areas, including diagnosis of several medical conditions, e.g., sleep disorders \cite{sleep_apnea}, brain tumours \cite{tumour} and epilepsy \cite{epilepsy} \cite{epilepsy2}. With the advancement of human-machine cooperation, smart sensors, Internet of Things (IoT) applications and services~\cite{pal2021internet} \cite{pal2018fine} \cite{sreedevi2022application}, and intelligent wearable devices~\cite{pal2020access}, industries try to synergise the intelligence of the brain and the ability of cutting-edge machine learning methods for BCI designs~\cite{hosseini2017optimized} \cite{chicaiza2021brain}. BCIs involve the brain to \textit{sense}, \textit{think}, and \textit{act} against signals by decoding information from neural activity~\cite{siegel2003sense}~\cite{yudhana2019recognizing}.

There is significant research in the area of EEG, exploring how the human brain works \cite{aggarwal2022review} \cite{zhang2021survey} \cite{mahendra2021classification} \cite{pal2021development}. It has been generalised to four different frequency bands of signals that occur in the brain: (i) delta, (ii) theta, (iii) alpha, and (iv) beta. \textit{Delta} waves are most commonly associated with deep sleep. \textit{Theta} waves often signify meditation and/or drowsiness. \textit{Alpha} waves are correlated with relaxation and eyes closed. Finally, \textit{beta} waves generally show high concentration and thinking \cite{eeg_band_separation}. Due to the difference between the varying frequency bands, a frequency representation of the brainwaves can be useful for BCI applications.

BCIs can be developed by analysing different representations and parts of EEG signals. For example, by looking at the blinking pattern of a user, and analysing the alpha wave band, a computer can be controlled. In order to solve more complex forms of BCIs, machine learning models are often built to give specific outputs for BCI tasks~\cite{rasheed2021review}. 
BCI is a powerful communication tool between users and systems. In other words, BCI bridges the human brain and the external world that typically do not depend upon information delivery methods. The function of BCI within a mechatronic system plays a fundamental role in its design, including its accuracy, speed (e.g., for real-time applications), and usability (e.g., learnability, and ease of use)~\cite{abdulkader2015brain}.
Mechatronic systems include but are not limited to quadcopter drones, aged-care assistant, autonomous vehicles, healthcare, etc. For instance, the EEG-mechatronic system interface can particularly be useful in medical applications, where a user may not be able to move their muscles to control a robot, but is able to control it with their mind. Fundamentally, mechatronic systems are beneficial to assisting people, but often with the help of some human interaction. Utilising BCIs can enable a more seamless control methodology.

Despite its various successes, BCI needs to overcome several challenges to gain more user acceptance and seamless integration with the newly discovered technologies. One of those challenges is how consumer-grade EEG devices can be used to control mechatronic systems. The motivation of this research is to develop a system with computational intelligence that can be utilised to create an interface between the human brain and a mechatronic system. We aim at utilising a consumer-grade EEG headset to develop a BCI, specifically with the interest of integrating it with a mechatronic system, e.g., a quad-copter drone or a robotic arm. Furthermore, unlike the present proposals, we aim to develop a system that enables online EEG data classification, allowing for live system control using the BCI. The major contributions of the paper can be summarised as follows:

\begin{itemize}
    \item We provide a detailed discussion on consumer-grade EEG headsets to develop a BCI. We utilise a dry EEG electrode headset with a computer feedback system for a motor-imagery BCI.
    \item We provide a detailed design of the system. We determine optimal machine learning models and their subsequent architecture for a mechatronic system interface.
    \item We provide complete proof of concept prototype implementation of the system. Using real-world participants, we enable a command-based BCI control methodology using the headset and a video game.
    \item We present detailed evaluation results based on various machine learning models and compare their performance analysis.
\end{itemize}

The rest of the paper is organised as follows. In Section~\ref{chap:LitReview}, we discuss the background and preliminary work in the area of EEG technology. In Section~\ref{chap:sys-dev}, we discuss the design and development of a system used for the experimentation. In Section~\ref{chap:experiment}, the experimental procedures are discussed. In Section~\ref{chap:results}, the results from the experiment are detailed. In Section~\ref{chap:discussion}, the results are analysed and discussed. Section ~\ref{chap:related-work} presents some related work. Finally, in Section~\ref{sec:Conclusion}, we conclude the paper and discuss potential future work. In Table~\ref{tab:notation}, we provide a list of abbreviations and their full forms.

\label{chap:abbreviations}

\begin{table}[ht]
    \footnotesize
    \caption {List of abbreviations used in the paper and their full name.}
    \label{tab:notation}    
    \centering
    \scalebox{1}{
    \begin{tabular}{p{2.5cm}p{8cm}}
    \hline
    
    {Abbreviations} & {{Full name}}  \\ [0.5ex] \hline 
    
    ADC & Analog to Digital Converter \\ 
    ANN & Artificial Neural Network\\
    BCI & Brain-Computer Interface\\
    CAD & Computer-Aided Design\\
    CNN & Convolutional Neural Network\\
    DNN & Deep Neural Network\\
    EEG & Electroencephalography\\
    EI & Experiment Investigator\\
    FFT & Fast-Fourier Transformation\\
    KNN & K-Nearest Neighbours\\
    LDA & Linear Discriminant Analysis\\
    RNN & Recurrent Neural Network\\
    VEP & Visually Evoked Potential\\
    
    \hline
    \end{tabular}
    }
    
    \end{table}

    


\section{Background and Preliminaries} 
\label{chap:LitReview}
\begin{figure}[ht]
    \centering
    \includegraphics[width=0.8\textwidth]{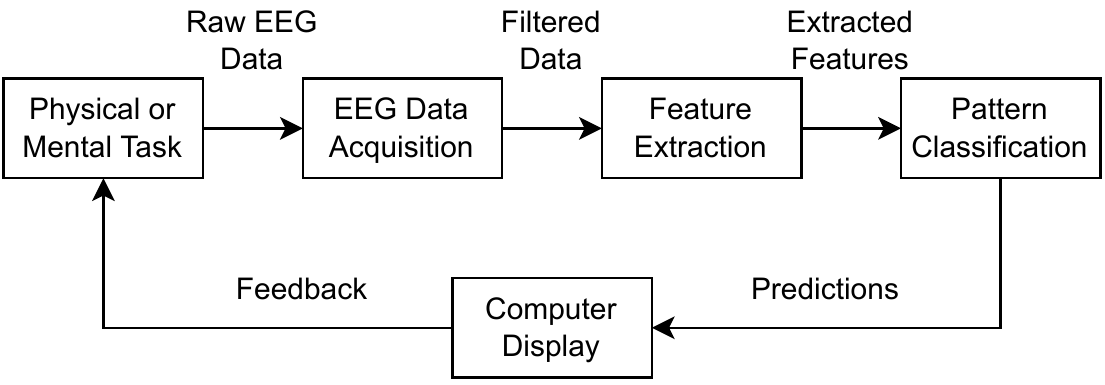}
    \caption{Components of an EEG BCI system.}
    \label{fig:bci-components}
\end{figure}

In this section, we provide a background discussion related to EEG BCI systems that are significant to our research, including the hardware, software, and BCI paradigms. In Fig.~\ref{fig:bci-components}, we illustrate the components of an EEG BCI system. 


\subsection{Hardware}
Hardware is one of the most fundamental parts of an EEG system. Important subcomponents of the hardware include the selected electrode used to read voltage potentials on the scalp, the electrical signal processing components, signal transmission, and the accessible data format of the data available to the researcher~\cite{gonzalez2021hardware}.


\subsubsection{Electrodes}
\begin{figure}[ht]
    \centering
    \includegraphics[width=0.7\textwidth]{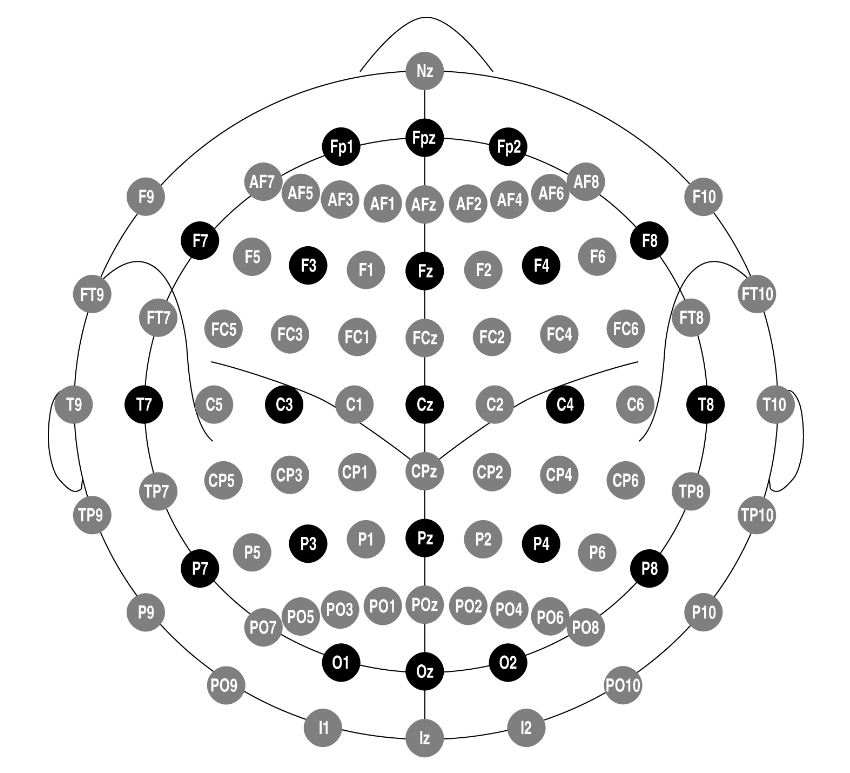}
    \caption[10-20 International System of Electrode Placement]{10-20 International System of Electrode Placement (Black) and 10-10 expanded system (Grey) \cite{10-20}.}
    \label{fig:10-20-system}
\end{figure}

EEG electrodes act as the interface between the scalp of an experimental participant and the electronic system of the controller/microprocessor which gathers the data. These electrodes are typically placed in positions with accordance to the 10-20 international system of electrode placement \cite{10-20}. The 10-20 international system can be seen in Fig.~\ref{fig:10-20-system} alongside the expanded 10-10 system.


\begin{figure}[ht]
    \centering
    \includegraphics[width=0.6\linewidth]{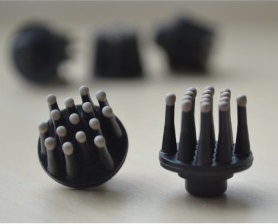}
    \caption{Dry pin-based EEG electrodes~\cite{GRUNDLEHNER2019223}.}
    \label{fig:dry-pins}
\end{figure}

Many types of electrodes exist, including but not limited to wet/saline based electrodes, pin-based dry electrodes, micro-spike dry electrodes, and flat-contact dry electrodes. Each of these electrodes have their own benefits and drawbacks.
Wet electrodes are more common in the medical industry due to their data consistency \cite{dry_revolution}. Conversely, they require a longer setup time compared to dry electrodes and sometimes require abrasive cleaning of the scalp \cite{novel_hybrid}. Wet electrodes also have data degradation issues due to the gel drying over time \cite{dry_noncontact}.
Dry electrodes generally consist of a conductive material contacting the scalp. These electrodes consist of varying types, from micro-spike electrodes \cite{micro_source}, flat-contact electrodes \cite{emotiv_website}, and pin-based electrodes \cite{dry_finger_3d}. The dry pin-based electrode will be used for this experiment as it is able to contact the scalp through hair, unlike the flat-contact \cite{novel_flat_dry, dry_revolution} and micro-spike electrodes \cite{micro_flexible}. These electrodes also have a consistent signal quality over time and require minimal setup~\cite{bristle_electrode}, unlike the micro-spike electrodes \cite{novel_mems_eeg}. This electrode can be seen in Fig.~\ref{fig:dry-pins}.

\subsubsection{Consumer-grade EEG headsets}
Since the invention of EEG technology, it has been primarily a tool used by the medical industry, but in 2009, NeuroSky released the MindFlex, an EEG controlled game designed for children. Since the release of the MindFlex, there has been an increased interest in consumer-grade EEG devices~\cite{katona2014evaluation}. Many new consumer-grade EEG devices have since been released, aiming to bring EEG technology outside the medical industry.

An important component of an EEG system is the headset. This is the component which converts the analog signals from the electrodes into digital figures which can be processed by a micro-controller. It also contains any analog filtering circuits or data transmission components~\cite{khosla2020comparative}.
OpenBCI is a company which developed open-source hardware consumer EEG devices. OpenBCI has two main boards they have designed for EEG applications: the \textit{Ganglion} and the \textit{Cyton}. The Ganglion has 4 channels with a 24-bit resolution\cite{openbci}. Research using this board is limited, but simple BCI tasks have been demonstrated with this board \cite{ganglion_movies}.
The Cyton board is an 8-channel (configurable to 16-channels with a daisy chain board) board with a 24-bit resolution\cite{openbci}. This board has been used for BCI applications including biometrics \cite{cyton_biometrics}, motor imagery \cite{cyton_feature_reduction} \cite{support_vector_channel}, auditory responses \cite{cyton_smarthome}, and need reporting for disable persons \cite{cytondaisy_hunger}.  
Further, the Ultracortex Mark IV Headset is a headset designed to be 3D printed using a hobby-level 3D printer. In Table \ref{tab:headset-comp}, we provide a comparison of the available EEG headsets in the market.

\begin{table}[ht]
\caption{Feature comparison of different consumer-grade EEG headsets.}
\label{tab:headset-comp}
\begin{center}
\scalebox{0.75}{%
\begin{tabular}{c | cccccc}
{Headset}                                                & {\begin{tabular}[c]{@{}c@{}}Electrode\\ Count\end{tabular}} & {\begin{tabular}[c]{@{}c@{}}Electrode\\ Type\end{tabular}} & {\begin{tabular}[c]{@{}c@{}}ADC\\ Res.\end{tabular}} & {\begin{tabular}[c]{@{}c@{}}Sample\\ Rate\end{tabular}} & {\begin{tabular}[c]{@{}c@{}}Transfer\\ Protocol\end{tabular}} & {\begin{tabular}[c]{@{}c@{}}Cost\\ (\$AU)\end{tabular}} \\ \hline
\begin{tabular}[c]{@{}c@{}}NeuroSky\\ MindFlex\end{tabular}     & 1                                                                  & \begin{tabular}[c]{@{}c@{}}Dry\\ (flat)\end{tabular}              & 12                                                          & 512                                                                    & \begin{tabular}[c]{@{}c@{}}Proprietary\\ 2.4 GHz\end{tabular}        & 102                                                            \\ \hline
\begin{tabular}[c]{@{}c@{}}NeuroSky\\ MindWave\end{tabular}     & 1                                                                  & \begin{tabular}[c]{@{}c@{}}Dry\\ (flat)\end{tabular}              & 12                                                          & 512                                                                    & BLE                                                                  & 228                                                            \\ \hline
\begin{tabular}[c]{@{}c@{}}Emotiv\\ EPOC\end{tabular}           & 14                                                                 & \begin{tabular}[c]{@{}c@{}}Saline\\ (wet)\end{tabular}            & \begin{tabular}[c]{@{}c@{}}14 or\\ 16\end{tabular}          & \begin{tabular}[c]{@{}c@{}}128 or\\ 256\end{tabular}                   & BLE                                                                  & 1098                                                           \\ \hline
\begin{tabular}[c]{@{}c@{}}Emotiv\\ Insight\end{tabular}        & 5                                                                  & \begin{tabular}[c]{@{}c@{}}Dry\\ (flat)\end{tabular}              & 14                                                          & 128                                                                    & BLE                                                                  & 387                                                            \\ \hline
\begin{tabular}[c]{@{}c@{}}Emotiv\\ EPOC Flex\end{tabular}      & 32                                                                 & \begin{tabular}[c]{@{}c@{}}Gel/Saline\\ (wet)\end{tabular}        & 14                                                          & 128                                                                    & BLE                                                                  & 2197                                                           \\ \hline
\begin{tabular}[c]{@{}c@{}}Emotiv\\ MN8\end{tabular}            & 4                                                                  & \begin{tabular}[c]{@{}c@{}}Dry\\ (flat)\end{tabular}              & 14                                                          & 128                                                                    & BLE                                                                  & ---                                                            \\ \hline
\begin{tabular}[c]{@{}c@{}}OpenBCI\\ Ganglion\end{tabular}      & 4                                                                  & Dry/Wet                                                           & 24                                                          & 200                                                                    & \begin{tabular}[c]{@{}c@{}}Proprietary\\ 2.4 GHz\end{tabular}        & 750                                                            \\ \hline
\begin{tabular}[c]{@{}c@{}}OpenBCI\\ Cyton\end{tabular}         & 8                                                                  & Dry/Wet                                                           & 24                                                          & 250                                                                    & BLE                                                                  & 1355                                                           \\ \hline
\begin{tabular}[c]{@{}c@{}}OpenBCI\\ Cyton + Daisy\end{tabular} & 16                                                                 & Dry/Wet                                                           & 24                                                          & 250                                                                    & BLE                                                                  & 2155                                                           \\ \hline
NextMind                                                        & 9                                                                  & \begin{tabular}[c]{@{}c@{}}Dry\\ (fingers)\end{tabular}           & ---                                                         & ---                                                                    & BLE                                                                  & 685                                                            \\ \hline
\begin{tabular}[c]{@{}c@{}}InteraXon\\ Muse\end{tabular}        & 2                                                                  & \begin{tabular}[c]{@{}c@{}}Dry\\ (flat)\end{tabular}              & \begin{tabular}[c]{@{}c@{}}12-\\ 16\end{tabular}            & \begin{tabular}[c]{@{}c@{}}220 or\\ 500\end{tabular}                   & BLE                                                                  & 325    \\                                                       
\hline
\end{tabular}
}
\end{center}
\end{table}


\subsection{Brain computer interfaces}
There are many different forms of BCIs which can be implemented with EEG headsets. Some of these paradigms required external stimulation, e.g., with visually evoked potentials (VEPs), while other paradigms only required actions taken by the user, such as with eyes-closed paradigms and motor imagery. 

\begin{figure}[ht]
\centering
\begin{subfigure}{0.5\linewidth}
    \centering
    \includegraphics[width=0.8\linewidth]{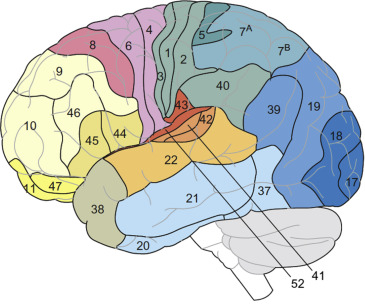}
    \caption{Brodmann's cortical map (lateral).}
    \label{brodmann-lat}
\end{subfigure}%
\begin{subfigure}{.5\linewidth}
    \centering
    \includegraphics[width=0.8\linewidth]{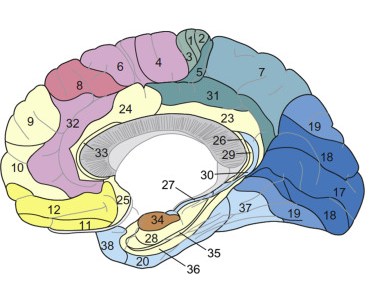}
    \caption{Brodmann's cortical map (medial).}
    \label{brodmann-medial}
\end{subfigure}
    \caption[Brodmann's cortical map]{Brodmann's cortical map~\cite{GAGE2018279}. Different colours represent different functions of the brain. For example, the light purple area 4 is responsible for motor movement, while the dark blue areas 17 and 18 are responsible for vision processing.}
    \label{fig:brodmann}
\end{figure}

It is well know that different regions of the brain perform different actions and respond differently to stimuli and thoughts \cite{brodmann_cortical_map}. For example, the motor cortex, located near the top of the head, is responsible for controlling muscle movements, while the visual cortex, near the back of the head, reacts to visual stimulus. A colour-coded cortical map can be seen in Figs. \ref{brodmann-lat} and \ref{brodmann-medial}. The relevant region and corresponding electrode placements for a BCI will depend heavily on the type of BCI paradigm used. 


Motor imagery is the process of analysing EEG signals to determine how the user is moving their muscles~\cite{motor_homunculus}. 
Motor imagery is popular for BCI tasks and has been used to control and play games \cite{motor_imagery_ball_game}. It has also been show to help improve the motor functions of those with limited motor function, e.g., paralysed people or those who have experienced a stroke \cite{BCI_boosts_stroke_recovery, BMI_stroke_rehab}. The motor cortex is located toward the top of the head, as shown in Fig. \ref{fig:brodmann}. Because of this, the ideal location for electrodes is around FCz, Cz, CPz, C3, C34, C5, and C6~\cite{motor_positions}.

\subsection{Data processing}
Proper data analysis and extraction is required for BCIs. While looking at EEG wave-forms and manually extracting features can sometimes be done, such as with the P300 or N170 paradigms, it can be complex to automate. There are many ways to automate this data analysis process, the most common of which is machine learning~\cite{BESSERVE2007}.

\subsubsection{Pre-processing/feature selection}
\label{sec:preprocessing}
Before analysing data from an EEG, it must first be preprocessed into a useful format. Preprocessing includes filtering, normalising, and formatting the data. Filtering is important to remove electrical noise from power lines, along with other environmental noise, and can be done with an Infinite Impulse Response (IIR) Filter \cite{7724548}, followed by a simple normalisation of data which should be formatted. This formatting can be of the form of a Fast-Fourier Transform (FFT), or an Independent Component Analysis (ICA).

\paragraph{Independent component analysis}
ICA is a method of separating separate independent samples from a given signal. An example of a problem ICA solves is ``cocktail party problem'' in which a microphone signal is analysed to isolate one person's voice from a room full of people speaking. This method of preprocessing is often used in offline EEG signal processing, such as in \cite{6952745} where EEG artefacts, such as eye-blinks, are omitted and SSVEPs are classified, and also in \cite{5334189} where noise and other artefacts are removed, and motor imagery is analysed. ICA is a computationally and time intensive algorithm and is often used for offline data preprocessing, but is not limited to such.

\paragraph{Wavelet transform}
A wavelet transform is another form of signal decomposition in which small wavelets are stretched and shifted to compose the full signal. The coefficients of these wavelets can be used for EEG BCI purposes, such as in one article \cite{bajaj_2020} which discusses the use of wavelets for pre-processing EEG signals and compares them to ICA, finding better results with the wavelet transform.

\paragraph{Fast-fourier transform}
FFT is an transformation algorithm which transforms a repeating signal from the time domain into the frequency domain, using a composition of sine and cosine waves. This allows for the analysing of signals with regards to how they are repeating rather than how they are changing with time. This transformation allows for the specific bands, alpha, beta, delta, and theta, to be analysed~\cite{eeg_band_separation}~\cite{eeg_sex}.

\subsubsection{Machine learning}
The specific type of Machine Learning relevant to this EEG BCIs is classification, in which a set of inputs is given and the model tells you what type of data has been given to it. The models discussed here are K-Nearest Neighbours, Linear Discriminant Analysis, and Artificial Neural Networks.

\paragraph{K-nearest neighbours}
K-Nearest Neighbours (KNN) is a classification algorithm which compares the input data with the training data and predicts the classification dependent on the closest k neighbours to the input. This algorithm requires all training data be loaded into memory, and hence can be quite resource intensive with large sets of training data. This algorithm has been used for BCI tasks, including recognition of emotion \cite{ml_many, emotion_knn} and motor imagery \cite{knn_motor}.

\paragraph{Linear discriminant analysis}
Linear Discriminant Analysis (LDA) is another classification machine learning algorithm which utilises the linear discriminant to split data into groups by reducing the system dimensions. In order to do this, a hyper-plane is created to project data onto in a way which maximises the distance between the centre of each classification while minimising the spread. LDA is often utilised for motor imagery \cite{lda_motor_imagery}, but has also been shown to work well for emotion classification \cite{lda_emotion}. 

\begin{figure}[ht]
    \centering
    \includegraphics[width=0.6\linewidth]{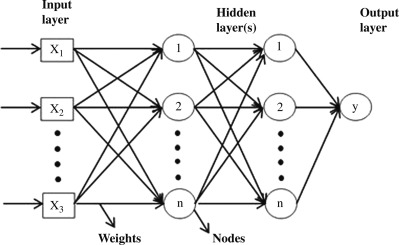}
    \caption[Artificial Neural Network architecture]{Artificial Neural Network (ANN) architecture~\cite{SAIRAMYA2019253}.}
    \label{fig:ann}
\end{figure}

\paragraph{Artificial neural networks}
Artificial Neural Networks (ANNs) are a machine learning model which tries to replicate the neural function of a brain \cite{sanchez2018artificial}. ANNs are often used for classification and regression. The generic layout of an ANN can be seen in Fig.~\ref{fig:ann}. With this architecture, an input layer is given the input data, which is multiplied by some weight with a bias added to it, and run through some activation function. This value is then sent to the nodes in the next layer, and so on.

Within ANNs, there is a subcategory called deep learning, in which an ANN has many layers, called Deep Neural Networks (DNNs)~\cite{dehghani2021deep}. Typically, DNNs require much more training data in order to get sufficient results, but are able to understand more complex systems. 

There are many types of deep learning, including but not limited to convolutional neural networks (CNNs) and recurrent neural networks (RNNs). Many papers discuss the use of different types of DNNs for classifying EEG data \cite{ann_recurrant, recurrent_nn_motor}. DNNs also allow for transfer learning, where a large dataset can be used to train the model, and a smaller dataset can be used to shape the model for the specific use case. This is done by locking some of the weights and biases in the nodes of the network and retraining on the new data. One article \cite{dnn_generic_transfer_learning} discusses the use of DNNs for general purpose EEG data transfer learning, allowing for EEG modelling between differing hardware, tasks, and subjects with a single model.

One article \cite{9492294} discusses the use of neural network transfer learning for emotion detection of users, helping get a deeper understanding of how transfer learning can be utilised in different studies for a better emotional understanding of experimental subjects. Another article \cite{8462115} discusses the successful use of transfer learning for motor imagery BCI applications, referencing EEG optical flow and image processing using common image classification CNNs such as VGG16 \cite{vgg16} and AlexNet \cite{alexnet}.

CNNs are a type of neural network in which has some convolutional layers. These convolutional layers are able to extract features from the input data, allowing for more optimised classification, most typically in the context of image processing. Several papers \cite{epilepsy_cnn, emotion_CNN} discuss the use of CNNs to analyse EEG signals for emotional classification and epileptic seizure detection, both of which could properly classify their targets with over an 80\% accuracy. One paper \cite{9495879} also discusses the use of a CNN for mental command purposes to control a robotic arm. This study discusses the ability to control the robotic arm with 8 separate commands with an accuracy of around 92\%.

\section{System design and development}
\label{chap:sys-dev}
In this section, we discuss the design and development of the EEG-mechatronic system interface. The following are discussed: the selection of the consumer-grade EEG headset, the placement of the EEG electrodes, data pre-processing methodology, and machine learning model development.

\subsection{Headset selection}
Many factors must be considered in order to determine which headset should be used in our experiment, e.g., the type of electrode, number of electrodes, sample rate, ADC resolution, transmission protocol, and cost. The OpenBCI Cyton board (without the Daisy Extension board) has been chosen as the EEG headset to be utilised for this research project. OpenBCI's available finger-based dry electrode is ideal for this application does not require a messy gel and is consistent over long periods of time. The relatively high ADC resolution in this headset is ideal to avoid a low signal to noise ratio. This headset has demonstrated success with BCI applications, suggesting its ease of use and reliability.

\subsection{Electrode placement}
\begin{figure}[ht]
\centering
\begin{subfigure}{0.4\linewidth}
    \centering
    \includegraphics[width=0.9\textwidth]{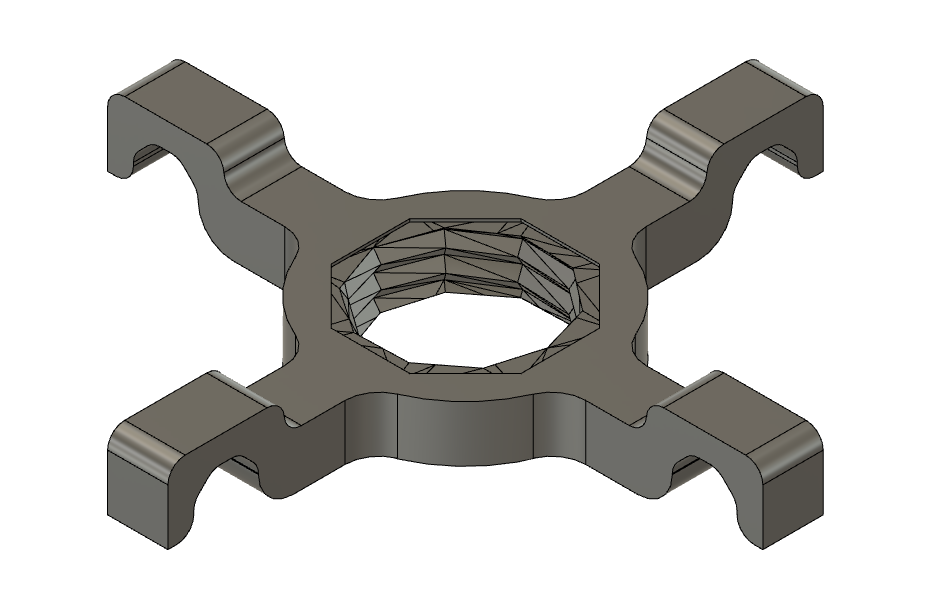}
    \caption{\small{Electrode adapter CAD render.}}
    \label{fig:electrode-adapter}
\end{subfigure}%
\begin{subfigure}{.55\linewidth}
    \centering
    \includegraphics[width=0.85\textwidth]{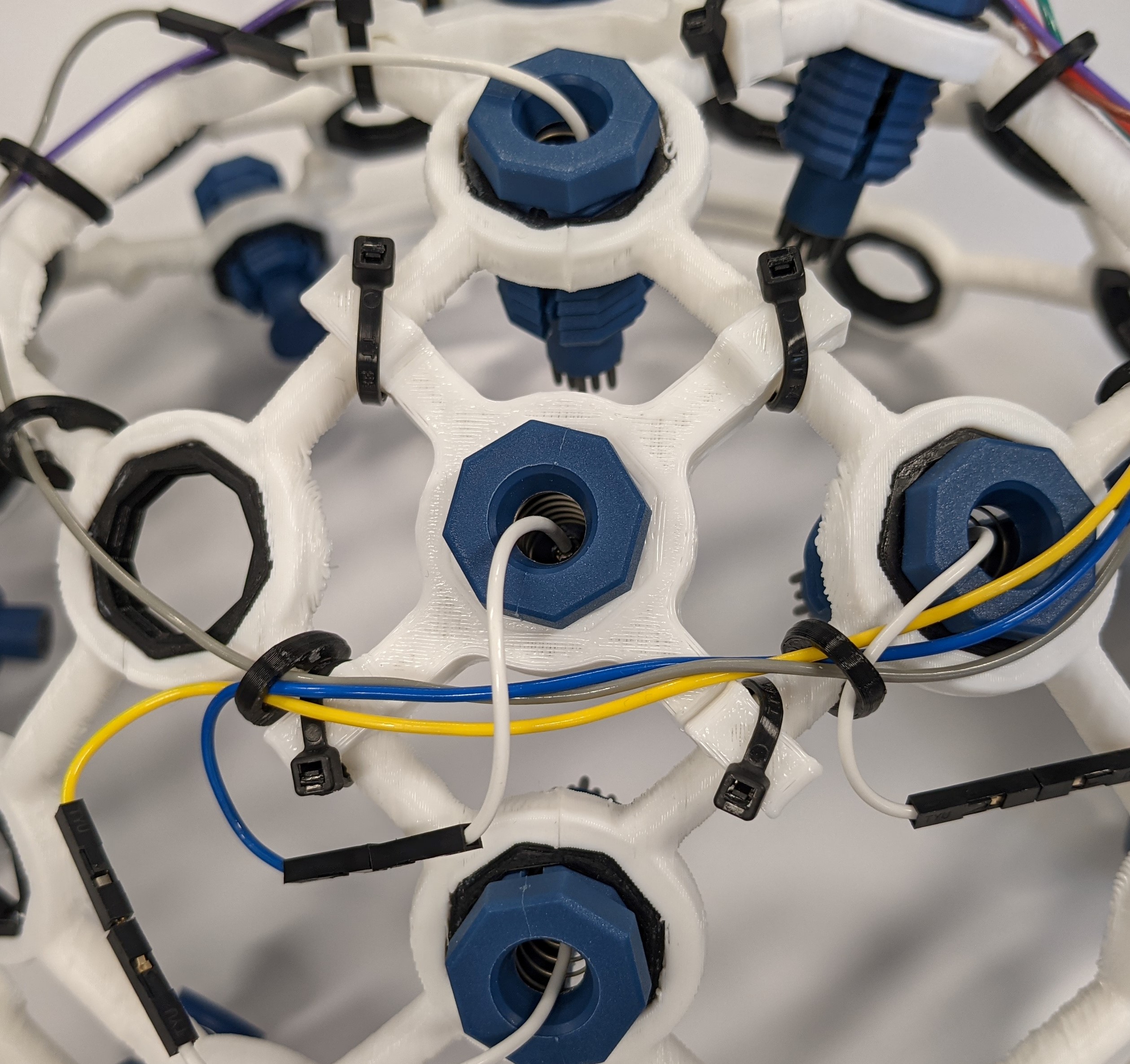}
    \caption{Electrode adapter on headset C1 position.}
    \label{fig:electrode-adapter-picture}
\end{subfigure}
    \caption{Electrodes C1 and C2 adapter as shown in (a) and (b).}
\end{figure}

The electrodes on the headset were be placed at 10-20 positions at the following locations: Cz, C1, C2, C3, C4, Cp1, Cp2, and Fpz. Because the Ultracortex Mark IV does not have electrode positions at C1 and C2, a 3D model was designed in the Fusion 360 Computer Aided Design (CAD) program to go between other electrodes on the headset. The model was created by creating a frame which could reach to the beams around positions C1 and C2. After this frame was designed, the ``MECH\-INSERT'' model provided by OpenBCI was merged with the frame to create the screw-in threads. The model was 3D printed, allowing for electrodes to be placed at C1 and C2, using cable ties to secure the parts to the headset. A render of this model can be seen in Fig.~\ref{fig:electrode-adapter}. This part can be seen mounted on the headset in Fig.~\ref{fig:electrode-adapter-picture}.



\begin{figure}[ht]
    \centering
    \includegraphics[width=0.8\textwidth]{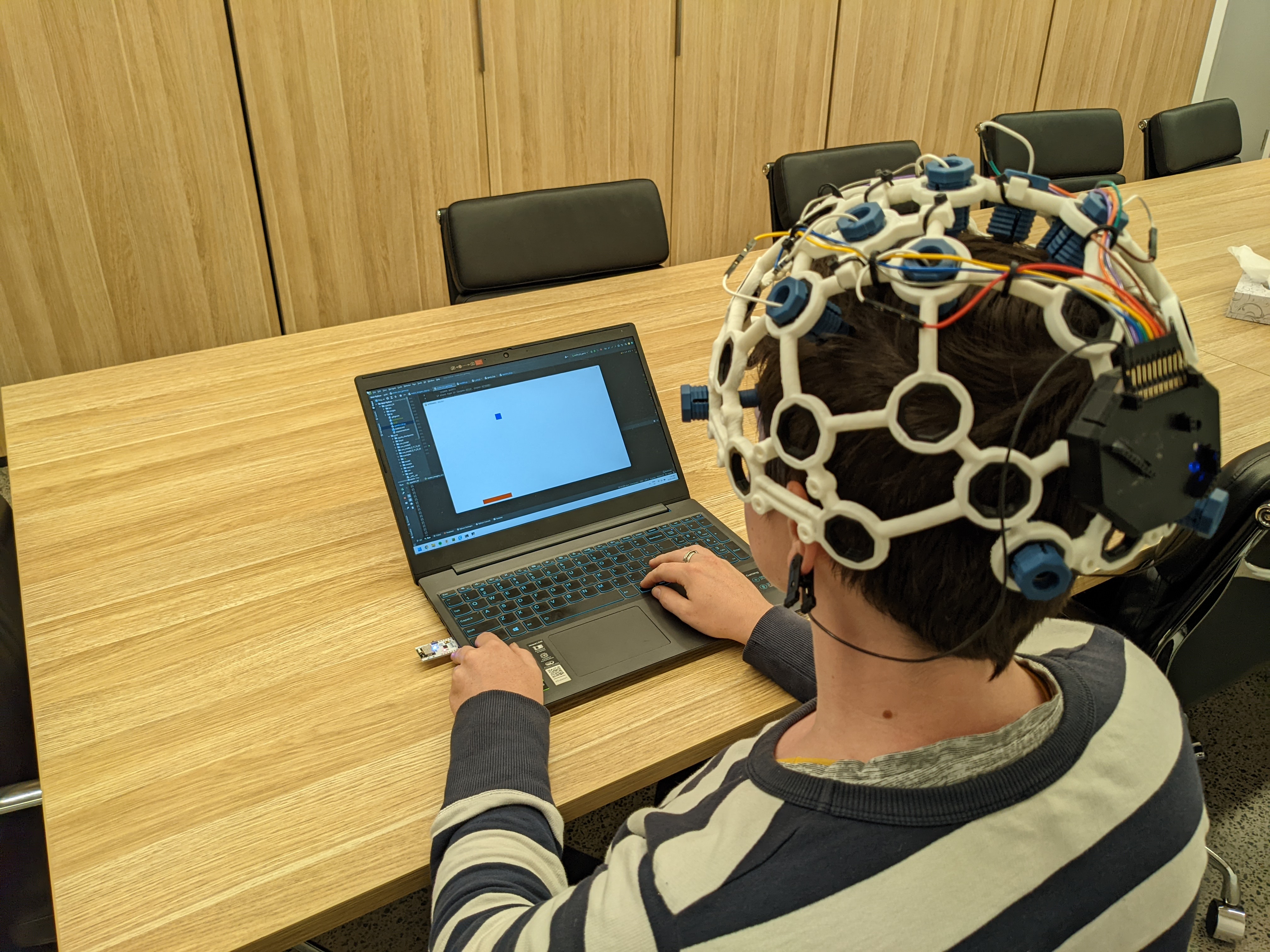}
    \caption{Participant wearing the headset and acquiring training data in our experiment.}
    \label{fig:participant-headset}
\end{figure}

\subsection{Data preprocessing}
After acquiring the EEG data, preprocessing should be done in order to clean the data and extract features. The process of filtering raw data and data balancing are discussed as follows:




\subsubsection{Digital data filtering}
\begin{figure}[ht]
    \centering
    \includegraphics[width=0.8\textwidth]{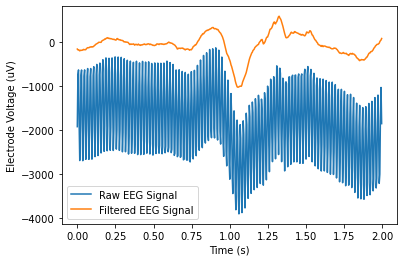}
    \caption{Filtered EEG data compared to unfiltered data.}
    \label{fig:filtered-data}
\end{figure}

EEG signals are very prone to environmental noise from varying electromagnetic sources, so several filters were applied to the signals.
In order to develop a digital filter for the EEG data, multiple filters are combined then quanitised. The filter used in this experiment is a combination of a low-pass filter to remove noise above the frequencies shown in EEG signals, a high-pass filter to remove low-frequency/DC noise, and a notch filter to remove noise from mains electricity. The resulting Infinite-Impulse Response (IIR) filter is of the form seen in Equation \ref{eq:complete}, where $x(t)$ is the input signal voltage at time $t$, $a-n$ and $b-n$ are the denominator and numerator coefficients respectively for the filter's discrete transfer function, $\Delta t$ is the time-step between subsequent signal acquisition, and $y(t)$ is the output signal voltage at time $t$. 
 A sample of EEG data filtered in this IIR filter can be seen in Fig.~\ref{fig:filtered-data}.

\begin{equation}
\label{eq:complete}
    y(t) = b-0x(t) + b-1x(t-\Delta t) + ... - a-1y(t-\Delta t) - ... - a-4y(t-4\Delta t)
\end{equation}

\subsubsection{Data balancing}
In order to have the best results, classification machine learning models generally work best with balanced data, meaning their training data has close to an equal number of samples for each classification. To ensure the data was balanced for the machine learning training, the count of each classification in the training samples was considered. The count of the classifier with the least number of samples was saved and that number of samples is randomly taken from each classifier to ensure the training data was balanced.

\subsection{Selection of machine learning models}
\label{sec:ml-models}
Different machine learning models (e.g., KNN, LDA, CNN) were tested throughout the development of this system, each with its own merits. In order to validate each machine learning model, EEG data, in an FFT format, was provided as the input for the model alongside the button pressed from the user, which was the desired model output. This data was split at a 70/30 ratio, where 70\% of the data was used to train the model and 30\% was utilised to test the model. 

\subsubsection{CNN architecture}
In order to find the most effective CNN architecture, many different hyper-parameters were tested. The architectures which were tested all started with a convolutional layer (50 filters, kernel size of 4, and sigmoid activation function), followed by a batch normalisation, and a max pooling (size of 2). These 3 layers were then added again for each desired number of convolutions $n$. After the convolutional layers were added, a flatten layer was added, followed by a dense layer with length $l$ and a sigmoid activation. Finally, a dense layer with 4 nodes was added for the outputs, with a softmax activation. This means there were two hyper-parameters that were tuned: the number of convolutions $n$ along with the dense layer length $l$.

\begin{figure}[ht]
    \centering
    \includegraphics[width=0.8\textwidth]{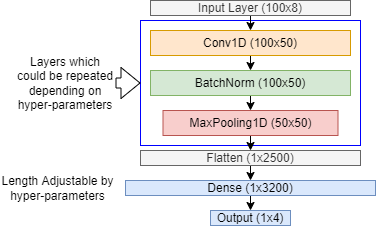}
    \caption[Architecture diagram of CNN]{Architecture diagram of CNN. The layers surrounded by a blue box can be optionally repeated to create more convolutional layers, batch normalisation layers, and max pooling layers, which would also impact the flatten layer's length. The length of the blue dense layer can also be adjusted with the hyper-parameters.}
    \label{fig:cnn-diagram}
\end{figure}

This machine learning model can be trained with 10\% of the training data, 100\% of the training data, and a specific participant's training data and the models can be saved. This is done with 1 convectional layer up to 4 convolutional layers, and with a dense layer length of 100, 200, 400, 800, 1600, and 3200, resulting in 24 unique CNN models for each training dataset, resulting in a total of 72 unique CNN models.

\section{Experimental procedures}
\label{chap:experiment}
In this section, we present the experimental procedures for the experiments involved in this  paper. The following are discussed: participant selection and demographics, experimental methodology, headset configuration, training data collection, machine learning training, and post-machine learning training procedures.



\subsection{Participants selection}
We aim to enable multiple participants to control a mechatronic system in a multi-week experiment. The training data collection portion of the experiment had a total of 19 participants, 6 female and 13 male. The participants ranged from 18 to 67 years old with a median of 23 years and a standard deviation of 12 years. Two of these participants had experience with EEG usage, while the rest were new to this area of research. The final validation portion of the experiment had 2 participants who both completed the training data portion of the experiment, who were both males aged 21 and 22 with experience with EEG usage.

\subsection{Experimental methodology}

We note that motor imagery is the selected BCI methodology for this experiment. The experiment consisted of two main portions: (i) the training data collection, and (ii) model validation, with a machine learning model training step in between the two. The training portion of the experiment consisted of a computer setup connected to the Cyton board via a bluetooth connection. The goal of this portion of the experiment was to collect motor-imagery training data to later consolidate and feed into a machine learning model. The machine learning model training portion of the experiment had the goal of combining all of the training data collected from all participants and training models to later determine which model best suits the given problem. The model validation portion of the experiment again consisted of a computer setup and the headset. The model validation read new data from the user wearing the headset and used transfer learning to re-train the machine learning models, which was followed by determining which model best suits the problem.

\subsubsection{Headset configuration}
\label{sec:headset-config}
\begin{figure}[ht]
\centering
\begin{subfigure}{0.45\linewidth}
    \centering
    \includegraphics[width=1\textwidth]{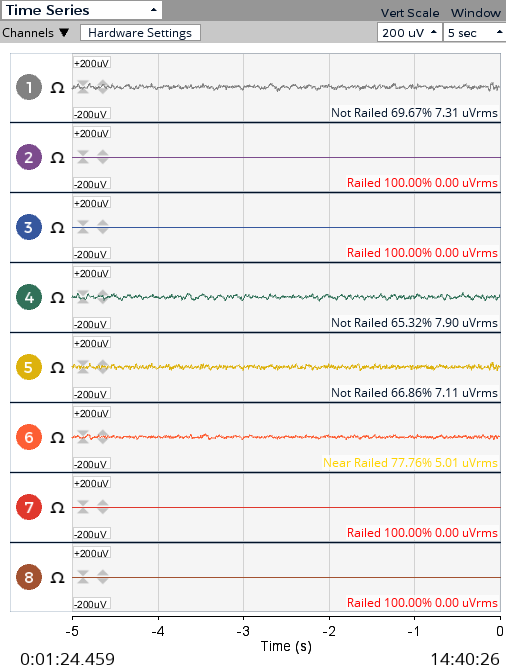}
    \caption{Poor electrode contact. Channels 2, 3, 7, and 8 are ``Railed 100\%''.}
\end{subfigure}%
\hspace{5mm}
\begin{subfigure}{.45\linewidth}
    \centering
    \includegraphics[width=1\textwidth]{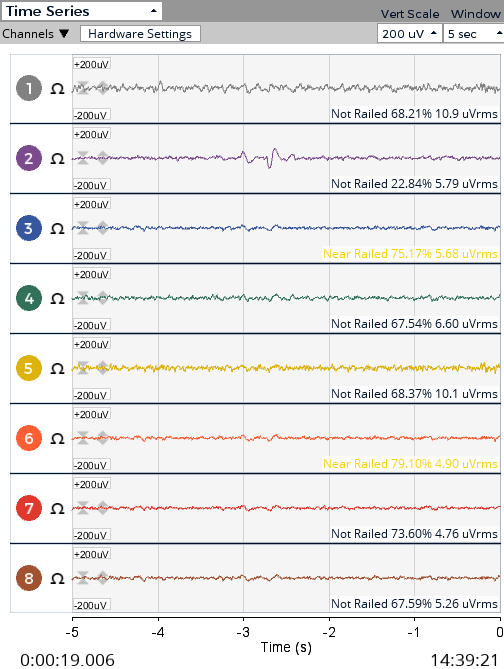}
    \caption{Good electrode contact. No channels are ``Railed 100\%''.}
\end{subfigure}
    \caption{OpenBCI GUI electrode contact.}
    \label{fig:electrode-contact-gui}
\end{figure}

In each 5-minute experimental session the participant had the headset placed on their head by the Experiment Investigator (EI). The electrode units were then be adjusted until each electrode was seated on the scalp of the participant. This process was done by running the ``OpenBCI GUI'' program to validate each channel is showing active brainwaves. Once the headset was properly seated, the EI followed the steps required to begin the data acquisition program. A comparison of poor electrode contact and good electrode contact can be seen in Fig. \ref{fig:electrode-contact-gui}, where poor contact is defined as when any of the 8 channels are ``Railed 100\%''. 

\subsubsection{Training data collection}
\label{sec:training-data}

In order to gather training data for the machine learning models, an experiment was made to provide information on the user's motor cortex when specific motor movements are performed. To get this data, the users were instructed to play a computer game which requires them to press keyboard keys with their hands. While the user was playing the video game, the EEG data was streamed to the computer, processed, and saved for future training sessions.

\begin{figure}[ht]
    \centering
    \fbox{\includegraphics[width=0.8\textwidth]{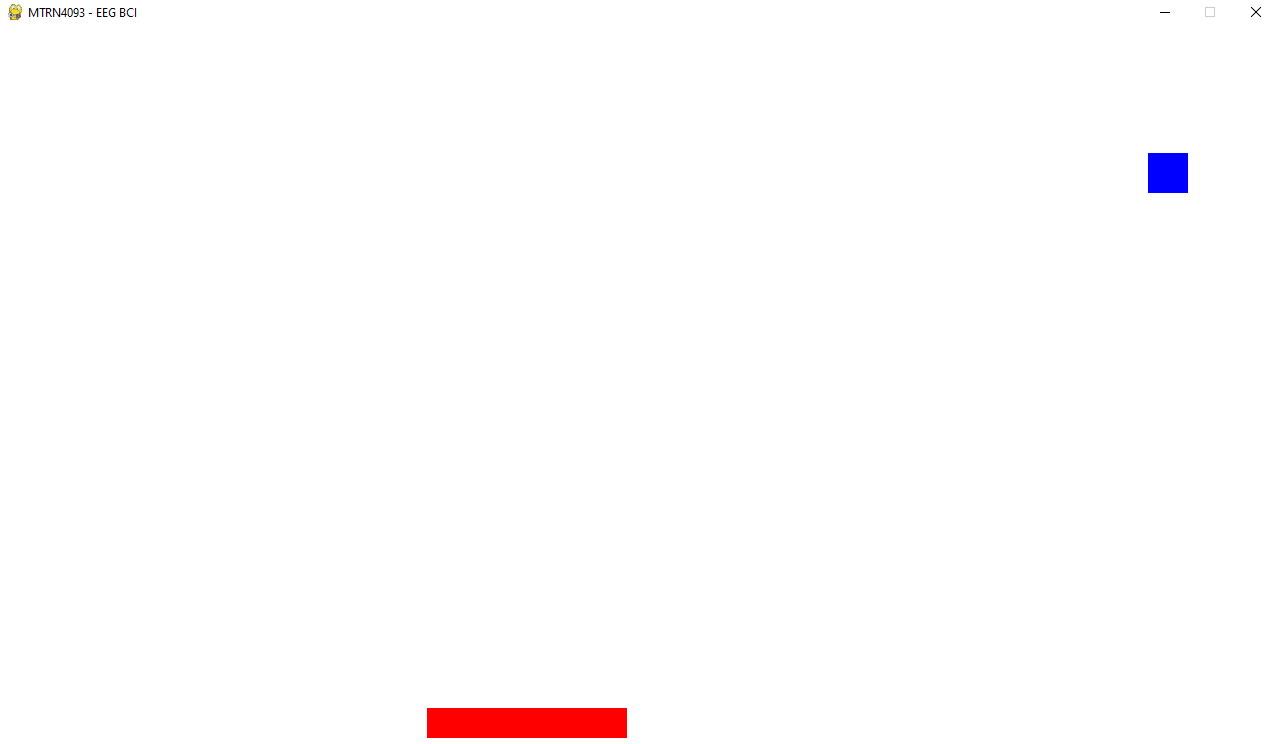}}
    \caption[BCI game screenshot]{Screenshot of the 2D game which the user plays. The blue square box falls downward continuously, and the user tries to move the red bar to catch the box. Once the blue box collides with the red bar or reaches the bottom of the screen, a new box appears randomly at the top of the screen.}
    \label{fig:bci-game-screenshot}
\end{figure}

Data was acquired while the participant was playing a 2-dimensional video game in which the user will move left or right by pressing the left and right `control' keys on a keyboard with fingers on their left and right hands respectively. A screenshot of this program can be seen in Fig. \ref{fig:bci-game-screenshot}. In this game, the participant was able to move a bar left and right, and was trying to catch a falling box. Once the box hit the bar, or the box hit the bottom of the screen, a new box appeared at a random point at the top of the screen. This game and data acquisition was run for 5 minutes for each participant.

After the first portion of the training data gathering, the training data was then fit into a machine learning model, with a training-test split of 70-30 respectively. The 70\% training data was used to train the machine learning model, and the remaining 30\% was used to provide a model accuracy to validate the model. This process generally only took 10-90 seconds, depending on the model used, thus the user would remain seated for the next portion of the experiment while this was occurring.

Once the machine learning model was built, a 1-minute demonstration of the machine learning model was conducted. During this period, the participant continued to press the buttons on the keyboard, but the output of the machine learning model controlled the game rather than the button presses. This demonstrated how well the model trained to the user with their current headset configuration.

\subsubsection{Data consolidation and machine learning training}
Once a total of 24 5-minute sessions were recorded, the data was be consolidated into one dataset for machine learning training. Two forms of consolidation take place: one which consolidated 100\% of each session of data, and another where only 10\% of each historical session was taken. Many variations of the machine learning models were trained in order to determine the most optimal architecture.

\subsubsection{Post machine learning training procedures}
Once the training-data collection portion of the experiment was complete and the machine learning models were built, the models were validated against live data. This was be done with the model validation portion of the experiment.

In this portion of the experiment, the headset first needed to be configured on a participant, as in Section \ref{sec:headset-config}. After configuring the headset, the participant played the game described in Section \ref{sec:training-data} for 30 seconds while the data was gathered. After gathering data, machine learning models with transfer learning capabilities were trained with the newly acquired data, while models without transfer learning capabilities were trained only with this 30 seconds of training. After training the model, a 30-second validation period of the game began, where the EEG data being fed into the machine learning model controlled the game. After each model was tested
a form discussing model responsiveness was completed by the participant.

The model accuracy, number of boxes caught, number of boxes caught in a row, and user responsiveness were compared between each model to determine which best suits this EEG-mechatronic system interface.

\section{Results and performance evaluation} 
\label{chap:results}

In this section, we present the achieved results and discuss the performance of our system. As stated in Section \ref{sec:ml-models}, many forms of machine learning classification algorithms were compared throughout this experiment, including KNN, LDA, and CNN.

\subsection{Results from KNN model}
This model was utilised in the beginnings of the development of this experiment. During testing, it was noticed that this model had overfitting issues. Whenever training data would be fit to the model, a very high testing accuracy would be seen (greater than 95\%), but when new data were provided, such as with live EEG data, the model would perform with significantly lower accuracy (around 40-45\%).

This overfitting was likely due to the data being relatively similar from sample to sample, as an FFT representation of data does not change significantly in a single time-step. In order to fix this overfitting issue, less of each data sample should be used (e.g. 1\% of the recorded data) and more sample data should be provided.

The KNN classification algorithm also requires all training data be loaded into the computer's memory for predicting values from new data. This means the more training data is loaded, the more system memory is required. With gigabytes of training data gathered from this experiment, it is a challenge to create a model with all of the training data. 

\subsubsection{KNN model validation}
During the machine learning validation portion of the experiment, only the data recorded in the data collection portion of the experiment (the 30 seconds prior to training) were used to train the model. Even with this limitation, the model appears to respond some amount to the EEG signals, see Table \ref{tab:knn-results}. Of the 10 trials of KNN validation between two participants, an average of 1.5 boxes were caught per session, with an average user rating of 2.2/5.

\begin{table}[ht]
\centering
\caption[KNN training accuracy]{KNN training accuracy, number of boxes caught during testing, and the user rating from the model.}
\label{tab:knn-results}
\begin{tabular}{c|cc|cc|cc}
Trial &
  \multicolumn{2}{c|}{\begin{tabular}[c]{@{}c@{}}Training\\ Accuracy\end{tabular}} &
  \multicolumn{2}{c|}{\begin{tabular}[c]{@{}c@{}}Boxes\\ Caught\end{tabular}} &
  \multicolumn{2}{c}{\begin{tabular}[c]{@{}c@{}}User\\ Rating\end{tabular}} \\ \hline
        & P2          & P6          & P2          & P6         & P2         & P6         \\ \hline
1       & 0.97        & 0.86        & 1           & 2          & 2          & 3          \\
2       & 0.95        & 0.88        & 0           & 0          & 1          & 1          \\
3       & 0.89        & 0.89        & 0           & 2          & 2          & 3          \\
4       & 0.88        & 0.88        & 2           & 3          & 2          & 3          \\
5       & 0.89        & 0.91        & 0           & 5          & 1          & 4          \\ \hline
Average & \multicolumn{2}{c|}{0.90} & \multicolumn{2}{c|}{1.5} & \multicolumn{2}{c}{2.2} \\
\hline
\end{tabular}
\end{table}

\subsection{Results from LDA model}
An LDA was also utilised in the beginnings of the development of this experiment. This classification algorithm resulted in a lower training accuracy compared to the KNN algorithm, but displayed higher validation accuracy with new data.

\subsubsection{LDA model validation}
Similarly to the KNN model validation, only the 30 seconds of training data were used for the LDA model validation because of the lack of transfer learning. During the machine learning validation portion of the experiment, participants caught an average of 1.3 boxes per session, and gave the responsiveness an average of 2.4/5, see Table \ref{tab:lda-results}.

\begin{table}[ht]
\centering
\caption[LDA training accuracy]{LDA training accuracy, number of boxes caught during testing, and the user rating from the model.}
\label{tab:lda-results}
\begin{tabular}{c|cc|cc|cc}
Trial &
  \multicolumn{2}{c|}{\begin{tabular}[c]{@{}c@{}}Training\\ Accuracy\end{tabular}} &
  \multicolumn{2}{c|}{\begin{tabular}[c]{@{}c@{}}Boxes\\ Caught\end{tabular}} &
  \multicolumn{2}{c}{\begin{tabular}[c]{@{}c@{}}User\\ Rating\end{tabular}} \\ \hline
  & P2   & P6   & P2 & P6 & P2 & P6 \\ \hline
1 & 0.56 & 0.39 & 0  & 0  & 1  & 2  \\
2 & 0.39 & 0.43 & 2  & 3  & 2  & 4  \\
3 & 0.44 & 0.39 & 2  & 1  & 3  & 3  \\
4 & 0.34 & 0.47 & 0  & 1  & 2  & 2  \\
5 & 0.53 & 0.53 & 2  & 2  & 2  & 3  \\ \hline
Average &
  \multicolumn{2}{c|}{0.45} &
  \multicolumn{2}{c|}{1.3} &
  \multicolumn{2}{c}{2.4} \\
  \hline
\end{tabular}
\end{table}

\subsection{Results from CNN model}
A CNN was also used as a classification algorithm in this experiment. In order to train the various CNN architectures, a high specification server would be required for various reasons discussed in Section \ref{sec:disc-harware}.


\subsubsection{CNN model with 10\% of training data}
\begin{figure}[ht]
    \centering
    \includegraphics[width=0.8\textwidth]{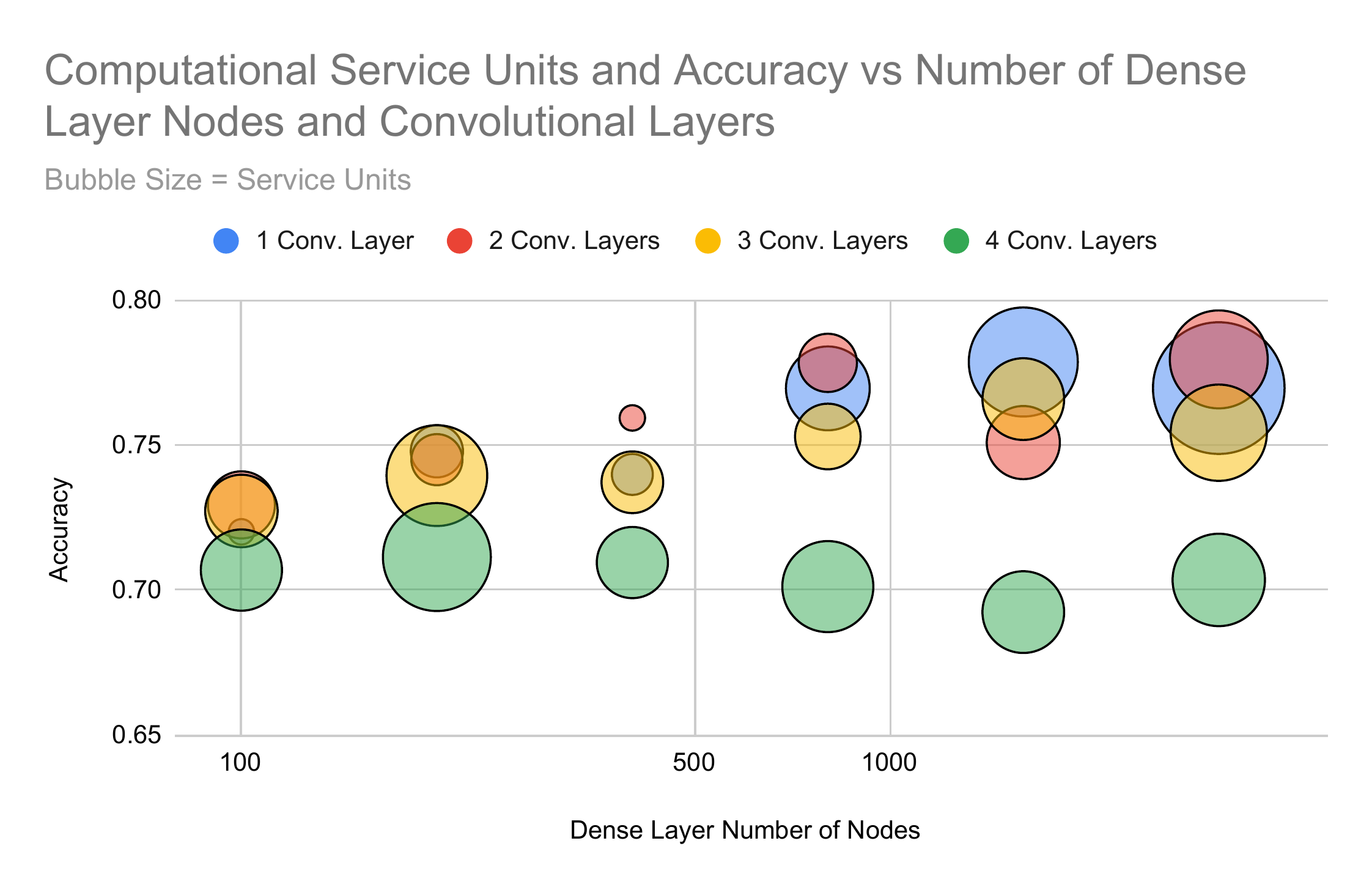}
    \caption[CNN 10\% training dataset accuracy]{Computational service units and accuracy vs number of dense layer nodes and convolutional layers with 10\% training dataset.}
    \label{fig:service-units-and-accuracy-vs-nodes-and-convs}
\end{figure}

24 different CNN models were trained with 10\% of the training data.
A plot of model accuracy  vs number of dense layer nodes can be seen in Fig.~\ref{fig:service-units-and-accuracy-vs-nodes-and-convs}. It can be seen that as the number of dense layer nodes increases from 100 to around 1600 the accuracy increases for 1, 2, and 3 convolutional layers, while the accuracy for 4 convolutions remained around a 70\% accuracy for all dense layer nodes. It can also be seen that for all dense layer lengths, 1 or 2 convolutional layers had the highest accuracy. Additionally 2 convolutional layers have the highest average accuracy and the lowest average service units, suggesting that 2 convolutional layers may be the best out of these CNN models.

\subsubsection{CNN model with full training data}
\begin{figure}[ht]
    \centering
    \includegraphics[width=0.8\textwidth]{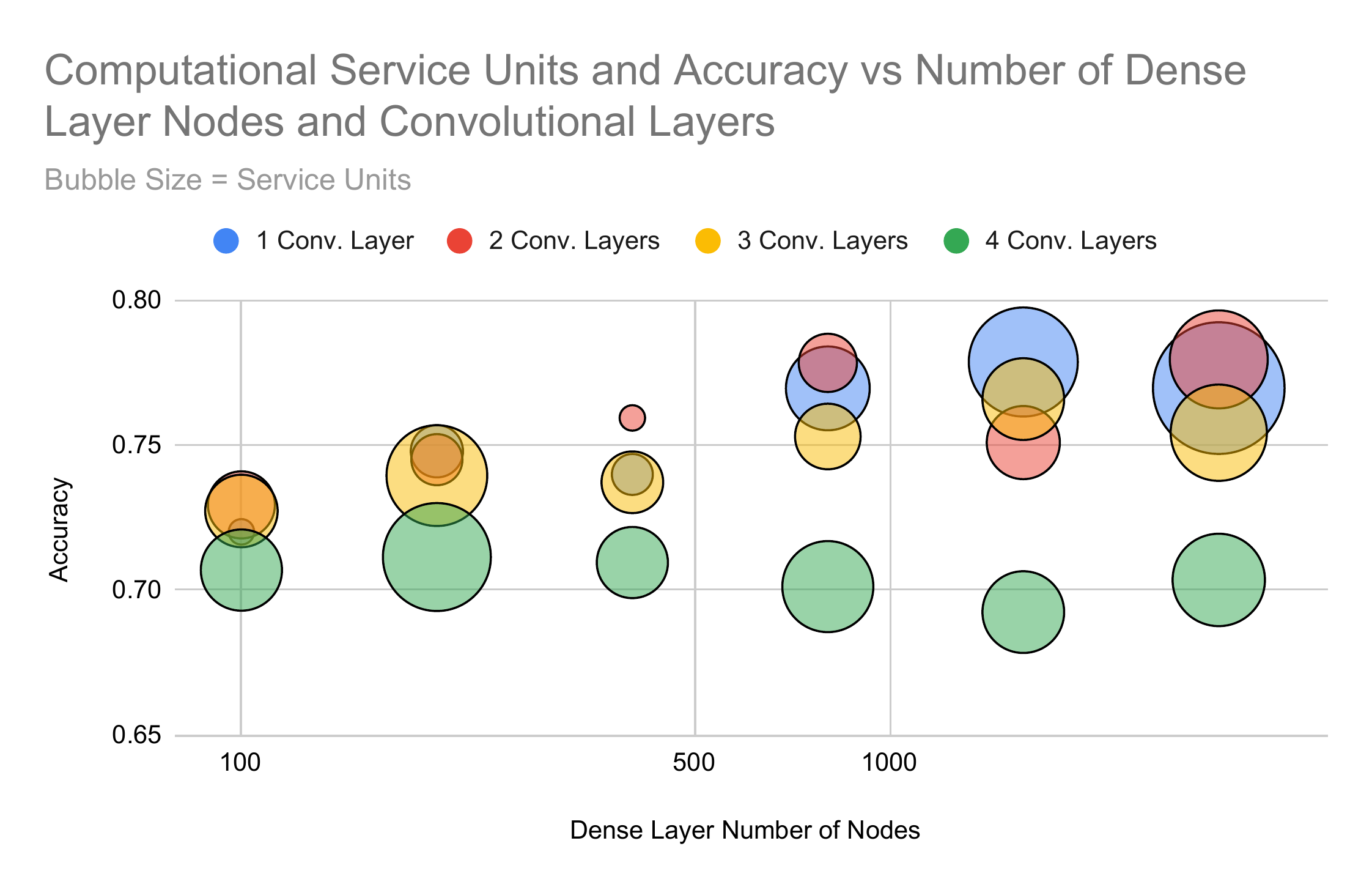}
    \caption[CNN full training dataset accuracy]{Computational service units and accuracy vs number of dense layer nodes and convolutional layers with full training dataset.}
    \label{fig:full-service-units-and-accuracy-vs-nodes-and-convs}
\end{figure}

The CNN architectures were also trained with the full training dataset. The accuracy of each model can be seen in Fig.~\ref{fig:full-service-units-and-accuracy-vs-nodes-and-convs}. In this, it can be seen that 1 and 2 convolutional layers again have the highest accuracy. A clear upward trajectory can be also seen for all number of convolutional layers up until around 800 dense layer nodes.  It can also be seen that the processing time increases with the number of convolutional layers.

\subsubsection{CNN model with specific user training data}
\begin{figure}[ht]
    \centering
    \includegraphics[width=0.8\textwidth]{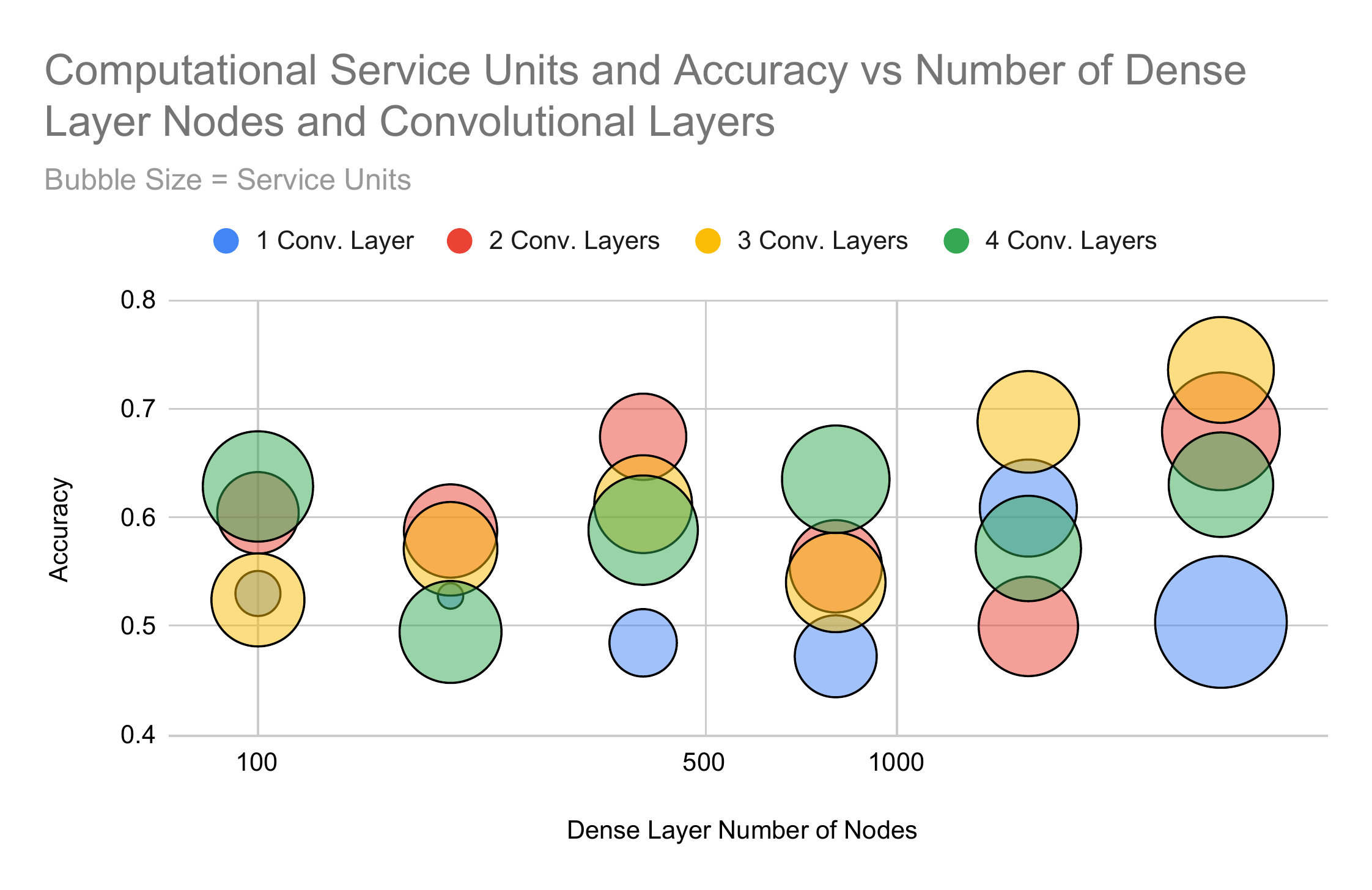}
    \caption[CNN user specific training dataset accuracy]{Computational service units and accuracy vs number of dense layer nodes and convolutional layers with participant 2 training dataset.}
    \label{fig:user-service-units-and-accuracy-vs-nodes-and-convs}
\end{figure}

In addition to training the CNN architectures with training data from all users, the models were also trained specifically using only the data from the participant with the most training data. The accuracy of each model can be seen in Fig.~\ref{fig:user-service-units-and-accuracy-vs-nodes-and-convs}.
In this, it can be seen that 2 and 3 convolutional layers again have the highest accuracy. No obvious trend is visible in this figure, other than a possible increase in 3 convolutional layers accuracy with an increase in number of dense layer nodes.

\subsubsection{CNN model validation}
Because there were 48 different CNN models, 48 validation experiments were required. The 10\% models were tested on two participants, while the full model was only tested on one participant.

The results for 10\% model can be seen in Table \ref{tab:cnn-validation-10}. It can be seen that 2 convolutional layers had the highest average user rating between the two subjects (2.6/5), and also had the highest average number of boxes caught (1.9 boxes). Within the models with 2 convolutional layers, the model with the highest average user rating was from the model with 100 dense layer nodes (4/5), while the highest average number of boxes caught was from the model with 3200 dense layer nodes (4 boxes).

\begin{table}[htb!]
     \caption[CNN model validation with 10\% training data]{CNN model validation for 1-4 convolutional layers and dense node numbers from 100-3200 with 10\% of training data.}
    \begin{subtable}[h]{0.475\textwidth}
        \centering
        \caption{1 convolutional layer}
        \scalebox{0.8}{%
        \begin{tabular}{c|cc|cc|cc}
        \begin{tabular}[c]{@{}c@{}}Dense\\ Number\end{tabular} &
          \multicolumn{2}{c|}{\begin{tabular}[c]{@{}c@{}}Training\\ Accuracy\end{tabular}} &
          \multicolumn{2}{c|}{\begin{tabular}[c]{@{}c@{}}Boxes\\ Caught\end{tabular}} &
          \multicolumn{2}{c}{\begin{tabular}[c]{@{}c@{}}User\\ Rating\end{tabular}} \\ \hline 
                & P2          & P6          & P2          & P6         & P2         & P6         \\ \hline
        100     & 0.94        & 0.96        & 0           & 1          & 2          & 1          \\
        200     & 0.81        & 0.87        & 4           & 2          & 4          & 3          \\
        400     & 0.85        & 0.83        & 1           & 1          & 2          & 2          \\
        800     & 0.49        & 0.76        & 1           & 0          & 1          & 1          \\
        1600    & 0.88        & 0.92        & 1           & 0          & 2          & 2          \\
        3200    & 0.73        & 0.89        & 4           & 1          & 4          & 2          \\ \hline
        Average & \multicolumn{2}{c|}{0.83} & \multicolumn{2}{c|}{1.3} & \multicolumn{2}{c}{2.2}  \\ \hline
        \end{tabular}
        }
    \end{subtable}
    \hfill
    \begin{subtable}[h]{0.475\textwidth}
        \centering
        \caption{2 convolutional layers}
        \scalebox{0.8}{%
        \begin{tabular}{c|cc|cc|cc}
        \begin{tabular}[c]{@{}c@{}}Dense\\ Number\end{tabular} &
          \multicolumn{2}{c|}{\begin{tabular}[c]{@{}c@{}}Training\\ Accuracy\end{tabular}} &
          \multicolumn{2}{c|}{\begin{tabular}[c]{@{}c@{}}Boxes\\ Caught\end{tabular}} &
          \multicolumn{2}{c}{\begin{tabular}[c]{@{}c@{}}User\\ Rating\end{tabular}} \\ \hline 
                & P2          & P6          & P2          & P6         & P2         & P6         \\ \hline
        100     & 0.81        & 0.85        & 4           & 0          & 5          & 3          \\
        200     & 0.86        & 0.84        & 5           & 0          & 3          & 2          \\
        400     & 0.88        & 0.96        & 2           & 0          & 2          & 2          \\
        800     & 0.78        & 0.84        & 1           & 0          & 1          & 2          \\
        1600    & 0.88        & 0.81        & 0           & 3          & 1          & 4          \\
        3200    & 0.74        & 0.85        & 6           & 2          & 3          & 3          \\ \hline
        Average & \multicolumn{2}{c|}{0.84} & \multicolumn{2}{c|}{1.9} & \multicolumn{2}{c}{2.6} \\ \hline
        \end{tabular}
        }
     \end{subtable}
     \vspace{5mm}
     \newline
    \begin{subtable}[h]{0.475\textwidth}
        \centering
        \caption{3 convolutional layers}
        \scalebox{0.8}{%
        \begin{tabular}{c|cc|cc|cc}
        \begin{tabular}[c]{@{}c@{}}Dense\\ Number\end{tabular} &
          \multicolumn{2}{c|}{\begin{tabular}[c]{@{}c@{}}Training\\ Accuracy\end{tabular}} &
          \multicolumn{2}{c|}{\begin{tabular}[c]{@{}c@{}}Boxes\\ Caught\end{tabular}} &
          \multicolumn{2}{c}{\begin{tabular}[c]{@{}c@{}}User\\ Rating\end{tabular}} \\ \hline 
                & P2          & P6          & P2          & P6         & P2         & P6         \\ \hline
        100     & 0.88        & 0.95        & 0           & 1          & 1          & 2          \\
        200     & 0.65        & 0.89        & 3           & 3          & 2          & 3          \\
        400     & 0.80        & 0.95        & 0           & 1          & 1          & 4          \\
        800     & 0.89        & 0.89        & 1           & 0          & 1          & 2          \\
        1600    & 0.85        & 0.87        & 0           & 1          & 2          & 3          \\
        3200    & 0.87        & 0.94        & 5           & 0          & 4          & 3          \\ \hline
        Average & \multicolumn{2}{c|}{0.87} & \multicolumn{2}{c|}{1.3} & \multicolumn{2}{c}{2.3} \\ \hline
        \end{tabular}
        }
    \end{subtable}
    \hfill
    \begin{subtable}[h]{0.475\textwidth}
        \centering
        \caption{4 convolutional layers}
        \scalebox{0.8}{%
        \begin{tabular}{c|cc|cc|cc}
        \begin{tabular}[c]{@{}c@{}}Dense\\ Number\end{tabular} &
          \multicolumn{2}{c|}{\begin{tabular}[c]{@{}c@{}}Training\\ Accuracy\end{tabular}} &
          \multicolumn{2}{c|}{\begin{tabular}[c]{@{}c@{}}Boxes\\ Caught\end{tabular}} &
          \multicolumn{2}{c}{\begin{tabular}[c]{@{}c@{}}User\\ Rating\end{tabular}} \\ \hline 
                & P2          & P6          & P2          & P6         & P2         & P6         \\ \hline
        100     & 0.93        & 0.93        & 0           & 4          & 1          & 4          \\
        200     & 0.97        & 0.94        & 0           & 1          & 2          & 2          \\
        400     & 0.80        & 0.93        & 2           & 0          & 2          & 3          \\
        800     & 0.89        & 0.81        & 3           & 2          & 3          & 3          \\
        1600    & 0.88        & 0.88        & 1           & 4          & 2          & 4          \\
        3200    & 0.87        & 0.93        & 1           & 1          & 1          & 2          \\ \hline
        Average & \multicolumn{2}{c|}{0.89} & \multicolumn{2}{c|}{1.2} & \multicolumn{2}{c}{1.8} \\ \hline
        \end{tabular}
        }
     \end{subtable}
     \label{tab:cnn-validation-10}
\end{table}

The results for full training-data model can be seen in Table \ref{tab:cnn-validation-full}. It should also be noted these models took significantly longer to train than the 10\% models. Because of this longer training time, and because the accuracy of each model and the number of boxes caught was lower than the 10\% models, these models were only tested on one subject. We see that 2 convolutional layers had the highest average user rating between the two subjects (2.6/5), and also had the highest average number of boxes caught (1.9 boxes). Within the models with 2 convolutional layers, the model with the highest average user rating was from the model with 100 dense layer nodes (4/5), while the highest average number of boxes caught was from the model with 3200 dense layer nodes (4 boxes).

\begin{table}[ht]
     \caption[CNN model validation with full training data]{CNN model validation for 1-4 convolutional layers and dense node numbers from 100-3200 for full training data.}
    \begin{subtable}[h]{0.475\textwidth}
        \centering
        \caption{1 convolutional layer}
        \scalebox{0.8}{%
        \begin{tabular}{c|c|c|c}
        \begin{tabular}[c]{@{}c@{}}Dense\\ Number\end{tabular} &
          \begin{tabular}[c]{@{}c@{}}Training\\ Accuracy\end{tabular} &
          \begin{tabular}[c]{@{}c@{}}Boxes\\ Caught\end{tabular} &
          \begin{tabular}[c]{@{}c@{}}User\\ Rating\end{tabular} \\ \hline 
        100     & 0.56 & 0    & 1    \\
        200     & 0.63 & 2    & 3    \\
        400     & 0.67 & 2    & 2    \\
        800     & 0.67 & 1    & 2    \\
        1600    & 0.70 & 3    & 3    \\
        3200    & 0.82 & 0    & 1    \\ \hline
        Average & 0.68 & 1.33 & 2.00 \\ \hline
        \end{tabular}
        }
    \end{subtable}
    \hfill
    \begin{subtable}[h]{0.475\textwidth}
        \centering
        \caption{2 convolutional layers}
        \scalebox{0.8}{%
        \begin{tabular}{c|c|c|c}
        \begin{tabular}[c]{@{}c@{}}Dense\\ Number\end{tabular} &
          \begin{tabular}[c]{@{}c@{}}Training\\ Accuracy\end{tabular} &
          \begin{tabular}[c]{@{}c@{}}Boxes\\ Caught\end{tabular} &
          \begin{tabular}[c]{@{}c@{}}User\\ Rating\end{tabular} \\ \hline 
        100     & 0.26 & 1    & 1    \\
        200     & 0.76 & 0    & 1    \\
        400     & 0.75 & 0    & 1    \\
        800     & 0.77 & 1    & 1    \\
        1600    & 0.67 & 0    & 1    \\
        3200    & 0.74 & 4    & 4    \\ \hline
        Average & 0.66 & 1.00 & 1.50 \\ \hline
        \end{tabular}
        }
     \end{subtable}
     \vspace{5mm}
     \newline
    \begin{subtable}[h]{0.475\textwidth}
        \centering
        \caption{3 convolutional layers}
        \scalebox{0.8}{%
        \begin{tabular}{c|c|c|c}
        \begin{tabular}[c]{@{}c@{}}Dense\\ Number\end{tabular} &
          \begin{tabular}[c]{@{}c@{}}Training\\ Accuracy\end{tabular} &
          \begin{tabular}[c]{@{}c@{}}Boxes\\ Caught\end{tabular} &
          \begin{tabular}[c]{@{}c@{}}User\\ Rating\end{tabular} \\ \hline 
        100     & 0.26 & 2    & 2    \\
        200     & 0.71 & 0    & 2    \\
        400     & 0.78 & 4    & 2    \\
        800     & 0.68 & 3    & 3    \\
        1600    & 0.88 & 0    & 1    \\
        3200    & 0.70 & 1    & 2    \\ \hline
        Average & 0.67 & 1.67 & 2.00 \\ \hline
        \end{tabular}
        }
    \end{subtable}
    \hfill
    \begin{subtable}[h]{0.475\textwidth}
        \centering
        \caption{4 convolutional layers}
        \scalebox{0.8}{%
        \begin{tabular}{c|c|c|c}
        \begin{tabular}[c]{@{}c@{}}Dense\\ Number\end{tabular} &
          \begin{tabular}[c]{@{}c@{}}Training\\ Accuracy\end{tabular} &
          \begin{tabular}[c]{@{}c@{}}Boxes\\ Caught\end{tabular} &
          \begin{tabular}[c]{@{}c@{}}User\\ Rating\end{tabular} \\ \hline 
        100     & 0.88 & 2    & 4    \\
        200     & 0.82 & 0    & 2    \\
        400     & 0.80 & 3    & 2    \\
        800     & 0.73 & 0    & 2    \\
        1600    & 0.84 & 0    & 2    \\
        3200    & 0.78 & 2    & 2    \\ \hline
        Average & 0.81 & 1.17 & 2.33 \\ \hline
        \end{tabular}
        }
     \end{subtable}
     \label{tab:cnn-validation-full}
\end{table}

The results for the user specific training-data model can be seen in Table \ref{tab:cnn-validation-user}. It can be seen that 3 convolutional layers had the highest average user rating by far with 3.3/5 along with the highest average number of boxes caught at 2.8 per session.

\begin{table}[ht]
     \caption[CNN model validation with user specific training data]{CNN model validation for 1-4 convolutional layers and dense node numbers from 100-3200 for participant 2's training data.}
    \begin{subtable}[h]{0.475\textwidth}
        \centering
        \caption{1 convolutional layer}
        \scalebox{0.8}{%
        \begin{tabular}{c|c|c|c}
        \begin{tabular}[c]{@{}c@{}}Dense\\ Length\end{tabular} &
          \begin{tabular}[c]{@{}c@{}}Training\\ Accuracy\end{tabular} &
          \begin{tabular}[c]{@{}c@{}}Boxes\\ Caught\end{tabular} &
          \begin{tabular}[c]{@{}c@{}}User\\ Rating\end{tabular} \\ \hline 
        100     & 0.57 & 2    & 2    \\
        200     & 0.51 & 1    & 1    \\
        400     & 0.62 & 0    & 1    \\
        800     & 0.57 & 0    & 1    \\
        1600    & 0.62 & 0    & 1    \\
        3200    & 0.80 & 1    & 1    \\ \hline
        Average & 0.61 & 0.67 & 0.67 \\ \hline
        \end{tabular}
        }
    \end{subtable}
    \hfill
    \begin{subtable}[h]{0.475\textwidth}
        \centering
        \caption{2 convolutional layers}
        \scalebox{0.8}{%
        \begin{tabular}{c|c|c|c}
        \begin{tabular}[c]{@{}c@{}}Dense\\ Length\end{tabular} &
          \begin{tabular}[c]{@{}c@{}}Training\\ Accuracy\end{tabular} &
          \begin{tabular}[c]{@{}c@{}}Boxes\\ Caught\end{tabular} &
          \begin{tabular}[c]{@{}c@{}}User\\ Rating\end{tabular} \\ \hline 
        100     & 0.56 & 3    & 2    \\
        200     & 0.66 & 1    & 3    \\
        400     & 0.53 & 1    & 2    \\
        800     & 0.81 & 2    & 2    \\
        1600    & 0.66 & 2    & 3    \\
        3200    & 0.69 & 1    & 1    \\ \hline
        Average & 0.65 & 1.67 & 2.17 \\ \hline
        \end{tabular}
        }
     \end{subtable}
     \vspace{5mm}
     \newline
    \begin{subtable}[h]{0.475\textwidth}
        \centering
        \caption{3 convolutional layers}
        \scalebox{0.8}{%
        \begin{tabular}{c|c|c|c}
        \begin{tabular}[c]{@{}c@{}}Dense\\ Length\end{tabular} &
          \begin{tabular}[c]{@{}c@{}}Training\\ Accuracy\end{tabular} &
          \begin{tabular}[c]{@{}c@{}}Boxes\\ Caught\end{tabular} &
          \begin{tabular}[c]{@{}c@{}}User\\ Rating\end{tabular} \\ \hline 
        100     & 0.60 & 2    & 4    \\
        200     & 0.65 & 5    & 5    \\
        400     & 0.69 & 1    & 2    \\
        800     & 0.78 & 5    & 3    \\
        1600    & 0.75 & 2    & 3    \\
        3200    & 0.73 & 2    & 3    \\ \hline
        Average & 0.70 & 2.83 & 3.33 \\ \hline
        \end{tabular}
        }
    \end{subtable}
    \hfill
    \begin{subtable}[h]{0.475\textwidth}
        \centering
        \caption{4 convolutional layers}
        \scalebox{0.8}{%
        \begin{tabular}{c|c|c|c}
        \begin{tabular}[c]{@{}c@{}}Dense\\ Length\end{tabular} &
          \begin{tabular}[c]{@{}c@{}}Training\\ Accuracy\end{tabular} &
          \begin{tabular}[c]{@{}c@{}}Boxes\\ Caught\end{tabular} &
          \begin{tabular}[c]{@{}c@{}}User\\ Rating\end{tabular} \\ \hline 
        100     & 0.70 & 2    & 5    \\
        200     & 0.63 & 0    & 1    \\
        400     & 0.72 & 1    & 1    \\
        800     & 0.81 & 2    & 2    \\
        1600    & 0.78 & 1    & 2    \\
        3200    & 0.73 & 1    & 2    \\ \hline
        Average & 0.73 & 1.17 & 2.17 \\ \hline
        \end{tabular}
        }
     \end{subtable}
     \label{tab:cnn-validation-user}
\end{table}

\section{Discussion and lessons learned}
\label{chap:discussion}
In this section, we analyse the results from  Section~\ref{chap:results}. We also provide the major findings of the results, and the limitations of our study. 


\subsection{Machine learning model analysis}
After performing the experiments, the results of KNN, LDA, and all configurations of the CNN should be compared and discussed. Each of these models will be analysed individually, with the final results compared between each configuration.

\subsubsection{Analysis of KNN model}
As seen in Table \ref{tab:knn-results}, the KNN machine learning model had an average training accuracy of 90\%, with an average number of boxes caught by the user of 1.5 per session and an average user rating of responsiveness of 2.2/5. It should also be noted the two participants had significantly different results when comparing the user ratings and number of boxes caught per session. During model validation, participant 2 was only able to catch an average of 0.6 boxes per session and gave the model an average responsiveness rating of 1.6/5, while participant 6 was able to catch an average of 2.4 boxes per session and gave an average rating of 2.8/5. Relative to the results of the other machine learning models, this model is respectable but not the best. One apparent advantage of this model is the lack of significant training data required to get some result, as only 30 seconds of training data were required to get the results discussed above.

\subsubsection{Analysis of LDA model}
As seen in Table \ref{tab:lda-results}, the LDA machine learning model had an average training accuracy of 45\%, with an average number of boxes caught by the user of 1.3 per session and an average user rating of responsiveness of 2.4/5. There was only a slight difference between the two participants' number of boxes caught with the LDA compared to the KNN, as participant 2 caught an average of 1.2 boxes per session and gave an average rating of 2/5, while participant 6 caught an average of 1.4 boxes per session and gave an average rating of 2.8. Similarly to the KNN model, the LDA is able to give some results with minimal training data, but is not able to utilise large amounts of training data.

\subsubsection{Analysis of CNN models}
From Table \ref{tab:cnn-validation-10}, it can be seen there is a larger difference between performance of the model between the two participants for most of the model configurations. Between all of the CNN architectures, participant 2 caught an average of 1.9 boxes and gave an average rating of 2.2, while participant 6 caught an average of 1.2 boxes but gave an average user rating of 2.6. Because participant 6 gave a higher average rating but caught fewer boxes on average, it is likely the system was responding reasonably well but was nearly missing catching the boxes. Conversely, for participant 2, the higher number of boxes caught but the lower average user rating suggest the system may have responded well some of the time but not for others, allowing for box catches while the system was responding well, but missing the boxes during the rest of the time.

It can also be seen that the CNN which was trained specifically for participant 2 had some of the highest number of boxes caught and user ratings from the models with 3 convolutional layers. This suggests CNNs which are specially trained for each user would perform better compared to the models which are trained with every user's data. This would be expected, as each individual's brain works differently and a one-size-fits-all solution would likely be more difficult than a tailored solution. Further studies are required to further validate this theory. It should also be noted that there does not appear to be a significant difference between the results dependent on the number of nodes in the parameterised dense layer.

\subsubsection{Summary of models}
From the machine learning results seen in Tables \ref{tab:knn-results}, \ref{tab:lda-results}, \ref{tab:cnn-validation-10}, \ref{tab:cnn-validation-full}, \ref{tab:cnn-validation-user}, it can be seen that overall, CNNs had the highest average number of boxes caught per session, suggesting it is best for this application. More specifically, a CNN with 3 convolutional layers which is specially trained for each participant of the experiment is ideal.

\subsection{Pre-processing}
This experiment was designed with an FFT in mind for the main preprocessing component. While FFT is common for BCI paradigms, other forms of preprocessing have also been shown to be viable for BCI applications, see Section \ref{sec:preprocessing}. 

Particularly, ICA and wavelet transform preprocessing methods should be considered. Because ICA can separate components from the signals, it could be utilised to determine when specific actions occur. Utilising the ICA could help separate and remove artefacts from the waveforms, such as from eye blinks, face twitches, or other unrelated movements. Separately from the ICA, a wavelet transform could also be used to extract features from EEG signals, as it has also been widely used within EEG BCI tasks.

In order to reduce overfitting of training data, another form of train-test split could be used. Within the scope of this experiment, a randomised train-test split was done. This means that after the training data was collected, a random 70\% was utilised for training and the remaining 30\% was used to test the model. This means that two consecutive datapoints could be utilised in the training and testing of the model, causing overfitting. If a temporal train-test split was done, where the first 70\% of the time-sorted data were used to train the model and the last 30\% of the data were used to test the model, overfitting could be reduced.

\subsection{Limitations}
\label{sec:disc-harware}
Due to the high density of the EEG data, several gigabytes of data were recorded. Because certain machine learning models require all training data to be loaded into the computer memory even after training, it is difficult to utilise the complete training data for specific models on a standard-specification laptop. Because of this limitation, training of the CNN models was performed on a supercomputer.

\section{Related Work}
\label{chap:related-work}
In this section, we provide a discussion of related works. Specifically, we intend to contrast similar papers with our own and discuss relevant similarities and our novel implementations. For example, 
proposal~\cite{Roy2022} discusses the use of a 118-channel EEG headset to implement a motor-imagery BCI system which can control a lower-limb exoskeleton. The system discussed was able to determine desired motions of the exoskeleton based on historical EEG data from the BCI competition III~\cite{bci_competition}. Because this data was from a competition and not from a live system, the results are likely to differ from a system in the real, live world.

In~\cite{Chaudhry_Khan_Palla_Singh_Deshmukh_2022}, authors discuss the use of steady-state visually evoked potentials (SSVEPs) to control a prosthetic arm using the EMOTIV EPOC +, a 16-channel wet EEG headset. The described system is able to utilise the SSVEP to detect commands from the EEG data using offline testing. This system is limited similarly to~\cite{Roy2022} because it was not implemented with online/live EEG data. 

Proposal~\cite{9719160} presents an analysis of EEG data to determine the cognitive state of airplane pilots. This would then be used to perform safety manoeuvres in the event of decreased cognitive activity or loss of consciousness. While this system is able to determine the cognitive state with high accuracy and could be used in conjunction with the research done in this article, it does not discuss an active control methodology for the controlled system.

Other proposals, e.g.,~\cite{8376886}, \cite{math9060606}, and \cite{9249937} discuss the use of an EEG motor imagery BCI system for a motor imagery BCI classification. These proposals discuss high machine learning classification accuracy, but do not discuss performance with online EEG experimentation or the corresponding activation accuracy, whereas the research done in our paper focuses in this specific area.

The use of transfer learning for EEG motor imagery tasks is discussed in additional proposals, e.g.,~\cite{KHADEMI2022105288} and \cite{ZHANG2021102144}. These proposals highlight the differences between participants and difficulties with transferring learning between them. This issue is addressed within our research as participants have machine learning models built with all participants data, along with models for individual participants as well. 

In summary, unlike the existing proposals, in our paper, we utilise online EEG data classification for motor imagery BCI. This allows for a true understanding of the performance of any machine learning models built for our system. Additionally in our paper, transfer learning is applied by using historical data of individuals, avoiding any issues with differences between participants.


\section{Conclusion and future work}
\label{sec:Conclusion}
In this paper, to answer the question `how can consumer-grade EEG devices be used to control mechatronic systems', an EEG BCI system was developed with the usage of the OpenBCI Cyton board and a user-interface running a game. Training data was gathered from real-world participants, and several machine learning models were built. After machine learning models were built, participants went through a model validation experiment where they played the game, gathered smaller amounts of training data, and applied transfer learning with this data to reconfigure the model to their specific head and placement of the headset. With this configuration, KNN, LDA, and CNN models all were able to control the system, with the CNN performing best out of the three in terms of user rated responsiveness. The outputs of the CNN model could further be used to control a mechatronic system. One way to improve design is to predict the intention to move muscles rather than their movements. Further, developing an intent-based motor imagery BCI would allow the system to be run while performing other motor actions without false positives, giving the system further versatility. However, these are separate research directions, and we leave these for future works.

\section*{References}
\bibliography{01-bib}

\end{document}